\documentclass[a4paper,cits]{JHEP3c}

\usepackage{epsfig,multicol,bbm,amsmath,amssymb}

\newcommand\fverb{\setbox\pippobox=\hbox\bgroup\verb}
\newcommand\fverbdo{\egroup\medskip\noindent%
            \fbox{\unhbox\pippobox}\ }
\newcommand\fverbit{\egroup\item[\fbox{\unhbox\pippobox}]}
\newbox\pippobox
\def\beq{\begin{equation}}
\def\eeq{\end{equation}}
\def\bea{\begin{eqnarray}}
\def\eea{\end{eqnarray}}
\def\beaa{\begin{eqnarray*}}
\def\eeaa{\end{eqnarray*}}

\def\putunder#1#2{\mathrel{
\setbox0=\hbox{#1}\setbox1=\hbox{\scriptsize #2} \dimen0=-0.5\wd0
\advance\dimen0 by -0.5\wd1 \dimen1=0.5\wd0 \advance\dimen1 by
-0.5\wd1
\hbox{\box0\kern\dimen0%
\vbox to 0pt {\hbox{\lower 0.7em \box1}\vss}%
\kern\dimen1} }}

\newcommand{\gsim}{\lower.7ex\hbox{$\;\stackrel{\textstyle>}{\sim}\;$}}
\newcommand{\lsim}{\lower.7ex\hbox{$\;\stackrel{\textstyle<}{\sim}\;$}}
\newcommand{\be}{\begin{equation}}
\newcommand{\e}{\end{equation}}

\title{de Sitter String Vacua from K\"ahler Uplifting}

\author{Alexander Westphal\\
    ISAS-SISSA and INFN, Via Beirut 2-4, I-34014 Trieste, Italy\\
    E-mail: \email{westphal@sissa.it}}

\preprint{SISSA-75/2006/EP\\November 29, 2006}  % OR: \preprint{Aaaa/Mm/Yy\\Aaa-aa/Nnnnnn}
\abstract{We present a new way to construct de Sitter vacua in
type IIB flux compactifications, in which the interplay of the
leading perturbative and non-perturbative effects stabilize all
moduli in dS vacua at parametrically large volume. Here, the
closed string fluxes fix the dilaton and the complex structure
moduli while the universal leading perturbative quantum correction
to the K\"ahler potential together with non-perturbative effects
stabilize the volume K\"ahler modulus in a $dS_4$-vacuum. Since
the quantum correction is known exactly and can be kept
parametrically small, this construction leads to calculable and
explicitly realized de Sitter vacua of string theory with
spontaneously broken supersymmetry.}


\begin{document}

%\maketitle  IS IGNORED %%%%%%%%%%%

\baselineskip=18pt

\section{Introduction}

Much of the recent progress in string theory is connected to the
discovery of an enormous number~\cite{BoussoP,kklt,sussk,dougl} of
stable and meta-stable 4d vacua in its low-energy effective
supergravities. The advent of this 'landscape'~\cite{sussk} of
isolated, moduli stabilizing minima marks considerable progress in
the formidable task of constructing realistic 4d string vacua. In
particular, one of the most pressing issues has been how to
stabilize the geometrical moduli of a compactification, and at the
same time address the tiny, positive cosmological constant that is
inferred from the present-day accelerated expansion of the
universe~\cite{WMAP3}. Recently, the use of closed string
background fluxes in string compactifications has been studied in
this context~\cite{GKP,CBachas,PolStrom,Michelson,
DasSeRa,TaylVaf,GVW,Vafa,Mayr,GSS,KlebStrass,Curio2,CKLT,HaaLou,BB,
DallAgata,KaScTr,silver,acharya,dlust}. Such flux
compactifications can stabilize the dilaton and the complex
structure moduli in type IIB string theory. Non-perturbative
effects such as the presence of $D$p-branes~\cite{Verl} and
gaugino condensation were then used by Kachru {\it et al}
(henceforth KKLT)~\cite{kklt} to stabilize the remaining K\"ahler
moduli in such type IIB flux compactifications (for related
earlier work in heterotic M-theory see~\cite{Curio1}).
Simultaneously these vacua allow for SUSY breaking and thus the
appearance of metastable $dS_4$-minima with a small positive
cosmological constant fine-tuned in discrete steps.
KKLT~\cite{kklt} used the SUSY breaking effects of an
$\overline{D3}$-brane to achieve this. Alternatively the effect of
D-terms on $D7$-branes has been considered in this
context~\cite{bkqu}.

Bearing in mind the importance of constructing 4d de Sitter string
vacua in a reliable way, one should note the problems of using
$\overline{D3}$-branes as uplifts for given volume-stabilizing AdS
minima. The SUSY breaking introduced by an $\overline{D3}$-brane
is explicit and the uplifting term it generates in the scalar
potential cannot be cast into the form of a 4d ${\cal N}=1$
supergravity analysis. Thus, the control that we have on possible
corrections in supergravity is lost once we use
$\overline{D3}$-branes for SUSY breaking. Replacing them by
D-terms driven by gauge fluxes on $D7$-branes~\cite{bkqu} is one
way to alleviate this problem because then the SUSY breaking is
only spontaneous (for a detailed study of the D-terms from
magnetized $D7$-branes see e.g.~\cite{JockersLouis}). In this case
the requirements of both 4d supergravity and the $U(1)$ gauge
invariance necessary for the appearance of a D-term place
consistency conditions on the implementation of a D-term (noted
in~\cite{bkqu}, and emphasized in~\cite{BDKP,Nilles2,DuVe,VZ}).
These conditions have not yet been met by any concrete stringy
realization of~\cite{bkqu}, where the proposal was made in the
context of KKLT. More recently, progress has been made in finding
ways of having a D-term uplift co-existing consistently with
non-perturbative K\"ahler moduli stabilization both in 4d
supergravity~\cite{VZ} and in more stringy
contexts~\cite{DuVe,Carlos,ChoiJeong,DuMamb,Luest,hebtrap}. Note,
that the implementation of a consistent D-term uplift becomes
considerably simplified~\cite{PW} in the case, that the
stabilization of the volume proceeds perturbatively through an
interplay of the leading $\alpha'$- and string loop corrections to
the K\"ahler potential~\cite{HG,BHK2} (related work in the context
of 5D supergravity appeared in~\cite{zurab}). Since the
calculation of the string loop correction (which has to be carried
out for each Calabi-Yau anew) is technically
challenging~\cite{BHK1}, it is not clear how far this dS vacua
from K\"ahler stabilization generalizes.

In view of this situation it becomes appealing to look for a
possibility of F-terms generating the dS vacuum. Four lines of
access have been studied here: Firstly, one may use the SUSY
breaking $(0,3)$ ISD $G_{(3)}$-fluxes of type IIB flux
compactifications to stabilize the complex structure moduli in
(metastable) minima of non-vanishing F-term~\cite{SaSilv}.
Secondly, one may use F-terms coming from the interactions of
hidden sector matter fields to uplift AdS minima towards de
Sitter~\cite{NillFterm} (for a related discussion of dS vacua in
M-theory see~\cite{Acharya2}). The third way considers strong
gauge dynamics. This can lead to the existence of metastable
F-term SUSY breaking minima along the lines of the ISS
proposal~\cite{ISS} which may then be used for uplifting
purposes~\cite{DuISS,AHKO}. The effective description of these
metastable vacua can be done in terms of generalized
O'Raifaertaigh models which has been studied in the context of
KKLT recently in~\cite{OKKLT1,OKKLT2}. These constructions provide
examples of a general analysis of 4d ${\cal N}=1$ supergravity
with an F-term uplifting sector which is separated from the moduli
sector~\cite{ReinoScrucca}. The fourth path, which will be pursued
in this paper consists of using the leading correction to the
K\"ahler potential given by an ${\cal
O}(\alpha'^3)$-correction~\cite{bbhl}. The $\alpha'$-correction
has recently been used to provide a realization of the simplest
KKLT $dS$-vacua with, however, either ${\cal O}(1)$
volume~\cite{Brama} or considerably large values of the
$\alpha'$-correction~\cite{West}. A combination of the
contributions to the scalar potential from D-branes and the
$\alpha'$-correction can also be used to stabilize the volume
modulus in a dS minimum~\cite{Bobk} (related discussions of the
effect of K\"ahler corrections on the stabilization of light
moduli appear in~\cite{deAlwis}).

The present work, which extends the results of Balasubramanian \&
Berglund~\cite{Brama}, will show that in type IIB flux
compactifications the interplay of the leading non-perturbative
contributions to the superpotential and the leading
$\alpha'$-correction to the K\"ahler potential can lead to volume
stabilization in a dS minimum at parametrically large volume while
keeping the value of the $\alpha'$-correction parametrically
small. To get the volume at large values it is necessary to have
the rank of the condensing gauge group living on a stack of
D7-branes wrapping the 4-cycle dual to the volume modulus at
larger values of ${\cal O}(30\ldots 100)$. We will then show that
this dS vacuum persists after including the flux stabilization of
the dilaton and the complex structure moduli. Thus, the setup will
be shown to lead to a full stabilization of all geometric moduli
in a parametrically controllable, metastable dS minimum which
breaks supersymmetry spontaneously through non-vanishing F-terms.
Since the vacuum energy of this dS vacuum is controlled by the
magnitude of the flux superpotential, the cosmological constant
can be fine-tuned by virtue of the large number of 3-cycles of
generic type IIB flux compactifications. Finally, we will show
that the quintic $C\mathbb{P}^4_{1,1,1,1,1}$ provides a reasonably
explicit example realizing this construction of dS vacua from
K\"ahler uplifting. This illustrates the fact, that due to the
universal nature of the leading $\alpha'$-correction (which
contributes on every Calabi-Yau with non-zero Euler number $\chi$)
these K\"ahler uplifted dS vacua should exist on all Calabi-Yau
3-folds with $\chi<0$ and arithmetic genus
$\chi(D)=1$~\cite{explicitmodels}. Here $D$ denotes the divisor of
the corresponding 4-fold in F-theory which projects back to the
4-cycle dual to the volume modulus. Of course, an appropriate
choice of fluxes is necessary to get the dilaton stabilized at
weak coupling and to tune the arising dS minimum to nearly zero
vacuum energy.

The paper is organized as follows. Section~\ref{Kahler} reviews
the leading quantum correction to the K\"ahler potential of type
IIB flux compactifications as well as the general argument that
the combined effect of the leading perturbative correction to the
K\"ahler potential and the leading non-perturbative contribution
to the superpotential can produce volume stabilization in a
metastable dS minimum. These results are then used in
Section~\ref{dS} to show that this structure can be extended to
shift the stabilized volume to (in principle) arbitrarily large
values. We proceed then in Section~\ref{STUdS} to demonstrate that
upon the inclusion of flux stabilization of the dilaton as well as
of the complex structure moduli we arrive at a full stabilization
of all geometric moduli in a true dS minimum which succeeds in
keeping the volume parametrically large and the K\"ahler
correction small. This is done using an explicit example given by
the quintic hypersurface $C\mathbb{P}^4_{1,1,1,1,1}$ providing
evidence that these K\"ahler uplifted metastable $dS$-vacua can be
explicitly realized in type IIB string theory, and are thus
expected to exist for all Calabi-Yau 3-folds with
$h^{2,1}>h^{1,1}=1$, at least. We extend these results to other
examples of Calabi-Yau 3-folds with $h^{1,1}=1$. Finally, we
discuss and summarize our results in the Conclusion.

\section{The leading $\alpha'$-correction in KKLT}\label{Kahler}

Our discussion will take place in the framework of type IIB string
theory compactified to 4d on orientifolded Calabi-Yau threefolds
in the presence of RR and NS-NS closed-string background fluxes
along the lines of~\cite{GKP}. Thus, it will prove to be useful to
recall some of the main results here. Non-zero background flux
quantized on the 3-cycles of the Calabi-Yau requires the presence
D3-branes or 4-cycle wrapping D7-brane to source the flux and
induces a non-trivial warpfactor in the internal dimensions. This
leads to a geometry - visualized in Fig.~\ref{throat} - where the
warped manifold is conformally Calabi-Yau. Under certain
conditions it develops a region warped into an approximate AdS
throat region which ends in the UV on the bulk of the Calabi-Yau
and is capped off smoothly in the IR by an appropriate analogue of
the Klebanov-Strassler solution.

\FIGURE[ht]{\epsfig{file=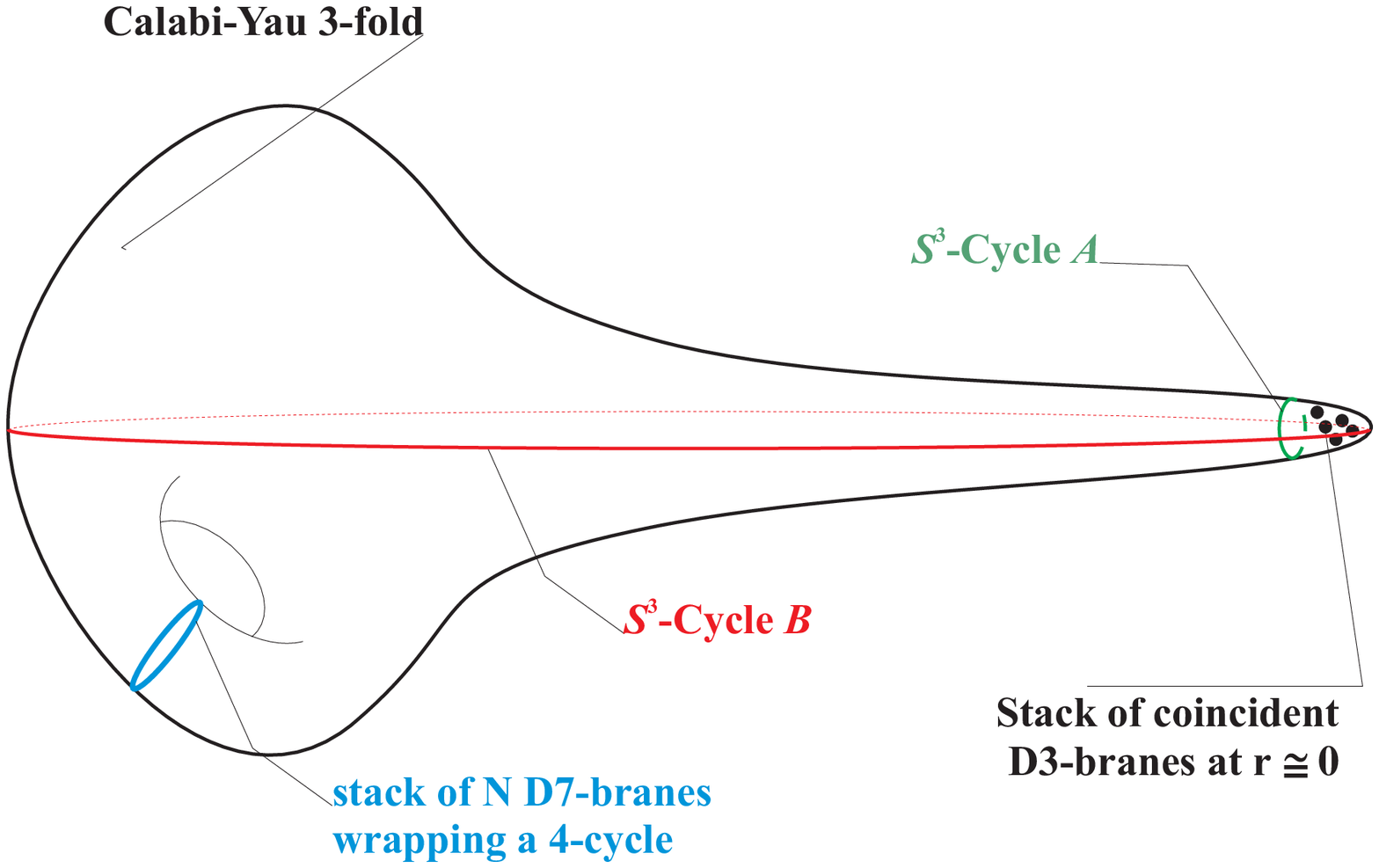,width=12cm} \caption{Flux induced
warped Calabi-Yau geometry in type IIB.}%
\label{throat}}

A very helpful property of these flux compactifications is the
fact that the equations of motion force the 3-form flux $G_{(3)}$
to be ISD and primitive in the $(1,2)$- or $(0,3)$-cohomology
classes of the Calabi-Yau. Simultaneously the back-reaction on the
geometry is confined to driving the non-trivial
warpfactor~\cite{GKP}. This we can see from the equations of
motion for the 5-form field strength $\tilde{F}_{(5)}$ and the
metric in the combination

\begin{small}\beq
\tilde{\nabla}^2(e^{4A}-\alpha)=e^{2A}\frac{|iG_{(3)}
-\ast_6\bar{G}_{(3)}|^2}{6\;\textrm{Re}\,S}+e^{-6A}\left|\partial(
e^{4A}-\alpha)\right|^2 +2\kappa_{10}^2
e^{2A}\left[\frac{1}{4}\,(T_m^m-T_{\mu}^{\mu})^{\textrm{loc}}-
\mu_3\overline{\rho_3^{\textrm{loc}}}\right]
\label{einbian}\eeq\end{small}

\noindent since the vanishing of the LHS requires the D-branes and
O-planes as the sources of flux to fulfill a pseudo-BPS condition
and $G_{(3)}$ to be ISD. This forces then $G_{(3)}$ to be of
$(1,2)$ or $(0,3)$ type and ensures that we stay in a geometry
which is conformally the same as the flux-less original
Calabi-Yau. The fluxes stabilize the type IIB axio-dilaton and the
complex structure moduli $U^\alpha$. This is encoded by the fact
that the fluxes generate a Gukov-Vafa-Witten type
superpotential~\cite{GVW} in the 4d effective supergravity
description \beq W_{\rm flux}=\frac{1}{2\pi}\,\int_{\rm
CY_3}G_{(3)}\wedge\Omega=:W_0\;\;.\label{fluxW}\eeq Then the fact
that $G_{(3)}$ being of $(1,2)$ or $(0,3)$ type keeps the
back-reaction confined to the warpfactor implies that the flux
superpotential can take values larger than ${\cal O}(1)$ while we
remain within the same supergravity solution. This will be very
helpful in the ensuing discussion.

Stabilization of the remaining K\"ahler moduli - which at
tree-level are no-scale - is then possible along the lines of
KKLT~\cite{kklt} by inclusion of the leading non-perturbative
effects in the superpotential like instantons from Euclidean
$D3$-branes wrapping 4-cycles of the Calabi-Yau or gaugino
condensation on 4-cycle wrapping stacked $D7$-branes. Thus, the
corresponding 4d ${\cal N}=1$ supergravity given by~\cite{kklt}
\bea K&=&-2\ln{\hat{\cal V}}-\ln(S+\bar{S})-\ln\left(-i\int_{\rm
CY_3}\bar{\Omega}\wedge\Omega\right)\nonumber\\
W&=&W_{\rm flux}+\sum_i A_i e^{-a_i T_i}\label{kkltsugra}\eea
generically manages to stabilize all geometric moduli and the
dilaton in a SUSY AdS minimum.

Here $\hat{\cal V}$ denotes the volume of the Calabi-Yau in
Einstein frame which is defined as $\hat{\cal
V}=\frac{1}{6}\,\kappa_{jkl}\,\hat{t}^j \hat{t}^k \hat{t}^l$ in
terms of the $h^{1,1}$ 2-cycle moduli $\hat{t}^j$ and the
intersection numbers $\kappa_{jkl}$. The 4-cycle K\"ahler moduli
$T_j$ are then defined by~\cite{bbhl,Bobk}
$T_j=\frac{1}{3}\,\partial_{\hat{t}^j}\hat{\cal
V}+i\int_{D_j}C_{(4)}=:T_{j,r}+i\tau_j$ where $D_j$ denotes the
divisor 4-cycle with volume ${\rm Re}\,T_j$ and $\hat{\cal V}$ is
thus a function of the $T_i$ defined implicitly through the
inverse of the former relations. For the following discussions we
will assume for simplicity the presence of just one K\"ahler
modulus $T$ measuring the overall volume $\hat{\cal V}$ as it is
realized, e.g., by the quintic example
$C\mathbb{P}^4_{1,1,1,1,1}$. Then we have $\hat{\cal
V}=\frac{\kappa}{6}\,\hat{t}^3=\gamma\,(T+\bar{T})^{3/2}$ where we
defined $\gamma=\frac{\sqrt{3}}{2\sqrt{\kappa}}$.

In addition to the leading non-perturbative effects we have now
the leading perturbative correction to the K\"ahler potential of
the K\"ahler moduli arising from the universal ${\cal
O}(\alpha'^3)$ $R^4$-correction to the 10d type IIB supergravity
action~\cite{bbhl,GreenSethi} \bea
S_{\textrm{IIB}}&=&-\frac{1}{2\kappa_{10}^2}\int
d^{10}x\sqrt{-g_{\textrm s}}\;e^{-2\phi}\left[R_{\textrm
s}+4\left(\partial\phi\right)^2+\left.\alpha^{\prime}\right.^3
\frac{\zeta(3)}{3\cdot
2^{11}}\;J_0\,+\,\ldots\,\right]\;\;.\label{aprim}\eea Here $J_0$
denotes the higher-derivative interaction {\small
\[J_0=\left(t^{M_1N_1\cdots M_4N_4}
t_{M_1^{\prime}N_1^{\prime}\cdots M_4^{\prime}N_4^{\prime}}
\hspace{-0.3ex}+\hspace{-0.5ex}\frac{1}{8}\;\epsilon^{AB
M_1N_1\cdots M_4N_4}\epsilon_{AB M_1^{\prime}N_1^{\prime}\cdots
M_4^{\prime}N_4^{\prime}}\right)\hspace{-1ex}
\left.R^{M_1^{\prime}N_1^{\prime}}\right._{M_1N_1}\cdots
\left.R^{M_4^{\prime}N_4^{\prime}}\right._{M_4N_4}\;,\]}and the
tensor $t$ is defined in~\cite{tensort}.  This generates a
corrected K\"ahler potential of the K\"ahler moduli~\cite{bbhl}
\bea K&=&-2\cdot \ln\left({\cal
\hat{V}}+\alpha'^3\frac{\hat{\xi}}{2}\right)\;,\;\;\;
\hat{\xi}=-\frac{\zeta(3)}{4\sqrt{2}(2\pi)^3}\;\cdot\chi\cdot(S+\bar{S})^{3/2}
=:\xi\cdot(S+\bar{S})^{3/2}\label{K1}\;\;.\eea Here ${\cal
\hat{V}}=\gamma\,(T+\bar{T})^{3/2}$ denotes the Calabi-Yau volume
in Einstein frame, $\chi$ is the Euler number of the Calabi-Yau
and from now on we set $\alpha'=1$.

The inclusion of this correction into the class of models defined
by eq.~\eqref{kkltsugra} has been shown to lead either to non-SUSY
AdS minima for the K\"ahler moduli at exponentially large
volumes~\cite{Brama2} or to non-SUSY Minkowski and dS minima at
small volume $\hat{\cal V}\sim 2$~\cite{Brama} without the need of
either $\overline{D3}$-branes or D-terms. We shall now summarize
the construction of the latter by Balasubramanian \&
Berglund~\cite{Brama} since their general argument for the
existence of dS vacua in the $\alpha'$-corrected theory forms the
starting point for the further discussion.

Prior to the inclusion of the above leading $\alpha'$-correction
we have a SUSY AdS minimum for all moduli $U^\alpha$, $S$ and $T$
given by $D_T W=0$ in the supergravity of eq.~\eqref{kkltsugra}.
Now, in the KKLT regime of $|W_0|\ll 1$, $W_0<0$ turning on the
correction by giving $\hat{\xi}$ some value still gives a solution
to the corrected $D_T W=0$ as long as the $\alpha'$-expansion
parameter $\hat{\xi}/(2\,\hat{{\cal V}})$ is small in the minimum.
Now increase $|W_0|$. Increasing $|W_0|$ will shift the solution
to the full $\alpha'$-corrected SUSY condition $D_T W=0$ to ever
smaller values of $T$. Hence, there is a value $|W_0|=W_{\rm
crit}$ where the solution to $D_T W=0$ will give $\hat{\cal
V}=\hat{\xi}$. At $\hat{\cal V}=\hat{\xi}$, the induced scalar
potential~\cite{bbhl,Brama} \beq
V_F(T)=e^K\,\left(K^{i\bar{j}}D_{T_i}W
\overline{D_{T_j}W}-3|W|^2\right)\label{VF}\eeq has a singularity.
Thus, for $|W_0|=W_{\rm crit}-\epsilon$ we have a SUSY AdS minimum
at $\hat{\cal V}>\hat{\xi}$ in the geometric region of moduli
space while for $|W_0|=W_{\rm crit}+\epsilon$ this SUSY stationary
point has passed through $\hat{\cal V}=\hat{\xi}$ towards
$\hat{\cal V}<\hat{\xi}$.

However, we have that $\lim_{\hat{\cal
V}\to\hat{\xi}^{+}0}V_F=+\infty$ and due to the dominance of the
perturbative $\alpha'$-correction over the non-perturbative
superpotential terms at large volume we have $V_F$ approaching
zero from above for $\hat{\cal V}\to\infty$.\footnote{This is true
for the case of $h^{1,1}=1$ discussed here, as well as for several
K\"ahler moduli if they are taken to be of same size. In the case
of several K\"ahler moduli with hierarchical values the
non-perturbative contribution to $W$ may dominate at large volume
and lead to non-SUSY AdS vacua at exponentially large volumes,
see~\cite{Brama2}.} Furthermore, the scalar potential
eq.~\eqref{VF} is a continuous function of $W_0$ for all
$\hat{\cal V}\neq\hat{\xi}$. In addition, for very large $|W_0|$
the non-perturbative contribution to $W$ becomes negligible
implying that then \beq V_F(T)=e^K\cdot
3\hat{\xi}\cdot\frac{\hat{\xi}^2+7\hat{\xi}\hat{\cal V}+\hat{\cal
V}^2}{(\hat{\cal V}-\hat{\xi})(\hat{\xi}+2\hat{\cal
V})^2}\,|W_0|^2\label{VFhugeW0}\eeq which is positive definite and
decreases monotonically from $\infty$ to $0$ for $\hat{\cal
V}>\hat{\xi}$. Together, this implies that after increasing
$|W_0|$ from $|W_0|=W_{\rm crit}-\epsilon$ to $|W_0|=W_{\rm
crit}+\epsilon$ there will be a non-supersymmetric AdS minimum for
$T$ at $\hat{\cal V}>\hat{\xi}$. Upon increasing $|W_0|$ further
this non-SUSY AdS minimum will eventually become a Minkowski and
subsequently a dS minimum before disappearing altogether. This
general argument shows that there is a regime where the
combination of the leading perturbative effects in $K$ and the
leading non-perturbative effects in $W$ lead to dS vacua without
the need for a D-term or an uplifting $\overline{D3}$-brane.

A toy example was given in~\cite{Brama} where the choice of $A=1$,
$a=2\pi/10$, $W_0=-1.7$ and $\hat{\xi}\approx 0.4$ (as it is the
case, e.g., for the quintic with $\chi=-200$ if we assume $S$
stabilized at ${\rm Re}\,S=1$) led to the existence of a dS
minimum for ${\rm Re}\,T$ at ${\rm Re}\,T\approx 5$ corresponding
to $\hat{\cal V}\approx 2$. The stabilization of $S$ and the
complex structure moduli was assumed there.

In the main part of this paper we will now show that, firstly,
certain scaling properties of the scalar potential allow us to
stabilize the physical volume $\hat{\cal V}$, measured in Einstein
frame, at values of ${\cal O}(10^2\ldots 10^3)$ by fixing the
dilaton at weak coupling $g_S=({\rm Re}\,S)^{-1}\sim 0.1$.
Secondly, the stabilization of the dilaton $S$ and the complex
structure moduli $U^\alpha$ (at least for the case of a single
$U$) can be done explicitly in the above context. Simultaneously,
all axionic directions are shown to get lifted as well.

\section{Parametrically controllable K\"ahler uplifting and dS vacua}\label{dS}

The starting point of the ensuing discussion is the fact that the
$\alpha'$-corrected K\"ahler potential eq.~\eqref{K1} leads to a
mixing of the volume and the dilaton in the resulting scalar
potential of the theory. Therefore, to begin with we shall have to
write down the full F-term scalar potential for the $S$-$T$-moduli
sector of the model defined by \bea K&=&-2\cdot \ln\left({\cal
\hat{V}}+\frac{\hat{\xi}}{2}\right)
\;\;,\;\hat{\cal V}=\gamma\,(T+\bar{T})^{3/2}\nonumber\\
W&=&W_0+A e^{-a T}\;\;.\label{dSsugra}\eea Here $a=2\pi/N$ denotes
the beta function of the $SU(N)$ gauge theory living on a stack of
$N$ $D7$-branes which undergoes gaugino condensation. We will need
the inverse of the K\"ahler metric $K_{a\bar{b}}$ with $a,b=S,T$
which can be found, e.g., in~\cite{Bobk,Brama2} and is given by
\beq K^{a\bar{b}}=\left(\begin{array}{ll} \gamma^{-4/3}
\frac{\sqrt[3]{\hat{\cal V}} (4 \hat{\cal V}^2+\hat{\xi} \hat{\cal
V}+4 \hat{\xi} ^2)}{12 (\hat{\cal V}-\hat{\xi} )}
 & -\frac{3 (\gamma^{-1}\,\hat{\cal V})^{2/3} \hat{\xi}  \text{Re}(S)}{2(\hat{\cal V}-\hat{\xi}) } \\
 -\frac{3 (\gamma^{-1}\,\hat{\cal V})^{2/3} \hat{\xi}  \text{Re}(S)}{2(\hat{\cal V}-\hat{\xi}) }
 & \frac{(4 \hat{\cal V}-\hat{\xi} ) \text{Re}(S)^2}{\hat{\cal V}-\hat{\xi} }
\end{array}\right)\;\;.\label{K1ijinv}\eeq The scalar potential
then reads \bea V_F(S,T)&=&e^K\,\left(K^{a\bar{b}}D_a
W\overline{D_b W}-3|W|^2\right)\nonumber\\
&=&e^K\,\left\{K^{T\bar{T}}[W_T\overline{W_T}+(W_T\cdot\overline{W
K_T}+c.c.)]\right.\nonumber\\
&&\;\left.+[K^{T\bar{S}}D_TW\overline{D_SW}+c.c]+K^{S\bar{S}}|D_SW|^2\right.\nonumber\\
&&\left.+3\hat{\xi}\,\frac{\hat{\xi}^2+7\hat{\xi}\hat{\cal
V}+\hat{\cal V}^2}{(\hat{\cal V}-\hat{\xi})(\hat{\xi}+2\hat{\cal
V})^2}\,|W|^2\right\}\label{VFST}\eea where in $K_a$ and $D_aW$
the use of the full corrected K\"ahler potential is implied. We
will now first assume that the dilaton is stabilized by the fluxes
in a supersymmetric minimum (this will be justified later on,
where we will see that the full solution for $S$ and $T$, in fact,
stabilizes $S$ close to the supersymmetric point and we will get
$F_S\ll F_T$ but $m_S\gg m_T$). Then the scalar potential for $T$
becomes~\cite{Brama} \beq V(T_r)=e^K\,\left\{K^{T\bar{T}}[a^2 A^2
e^{-2 a T_r}+(-a A e^{-a T_r}\overline{W
K_T}+c.c)]+3\hat{\xi}\,\frac{\hat{\xi}^2+7\hat{\xi}\hat{\cal
V}+\hat{\cal V}^2}{(\hat{\cal V}-\hat{\xi})(\hat{\xi}+2\hat{\cal
V})^2}\,|W|^2\right\}\;\;.\label{VFT}\eeq Here we used that the
$T$-axion is stabilized at $\tau=0$ the same way as in the
original KKLT construction since the perturbative correction to
$K$ does not depend on $\tau$.

Plugging in here the values $A=1$, $a=2\pi/10$, $W_0=-1.7$, and
$\hat{\xi}\approx 0.4$ would then reproduce the dS minimum at
$\hat{\cal V}\approx 2$ of~\cite{Brama} if we assume ${\rm
Re}\,S=1$.

Now, let us note that eq.~\eqref{VFT} possesses a scaling property
of the following kind: under a rescaling \beq N\to \lambda
N\;,\;a=\frac{2\pi}{N}\to\lambda^{-1}a\;,\;T_r\to\lambda
T_r\;,\;\hat{\xi}\to\lambda^{3/2}\hat{\xi}\label{scale1}\eeq the
scalar potential scales as \[V_F\to\lambda^{-3} V_F\] while its
shape remains unchanged up to the fact that the transformation
eq~\eqref{scale1} stretches it along the $X$-axis.

This scaling behavior allows us to conclude that by choosing a
larger gauge group for the gaugino condensate\footnote{This
amounts to a choice of flux which via the tadpole conditions
determines the number of $D7$-branes stacked on the 4-cycle.} in
$W$ and rescaling $T_r$ and $\hat{\xi}$ appropriately we can get
dS minima for $T$ at parametrically large volumes $\hat{\cal
V}={\cal O}(100\ldots 1000)$. To give an example, let us take
$W_0\approx-1.7$, as before, but for the other
parameters~\footnote{This choice, although at the upper limit of
typical values of $N$, seems plausible as ranks $N$ of ${\cal
O}(30)$ have been discussed in the context of the
$C\mathbb{P}^4_{1,1,1,6,9}$-model with $h^{1,1}=2$
in~\cite{explicitmodels}. In any case, also $N=30$ would give a
viable model where we would get $T\sim 15$ and $\hat{\cal V}\sim
60$ (see below).} $a=2\pi/100$ and $\hat{\xi}\approx 79.8$ which
corresponds to a scaling with $\lambda=10$. In addition, let again
$\kappa=5$ as realized later on for $C\mathbb{P}^4_{1,1,1,1,1}$.

This choice of numbers then stabilizes $T$ in a dS minimum at
parametrically large (compared to the string scale given by
$\sqrt{\alpha'}$) volume. For the parameters chosen we get ${\rm
Re}\,T\approx 49$ which corresponds to $\hat{\cal V}\approx 376$.

\FIGURE[ht]{\epsfig{file=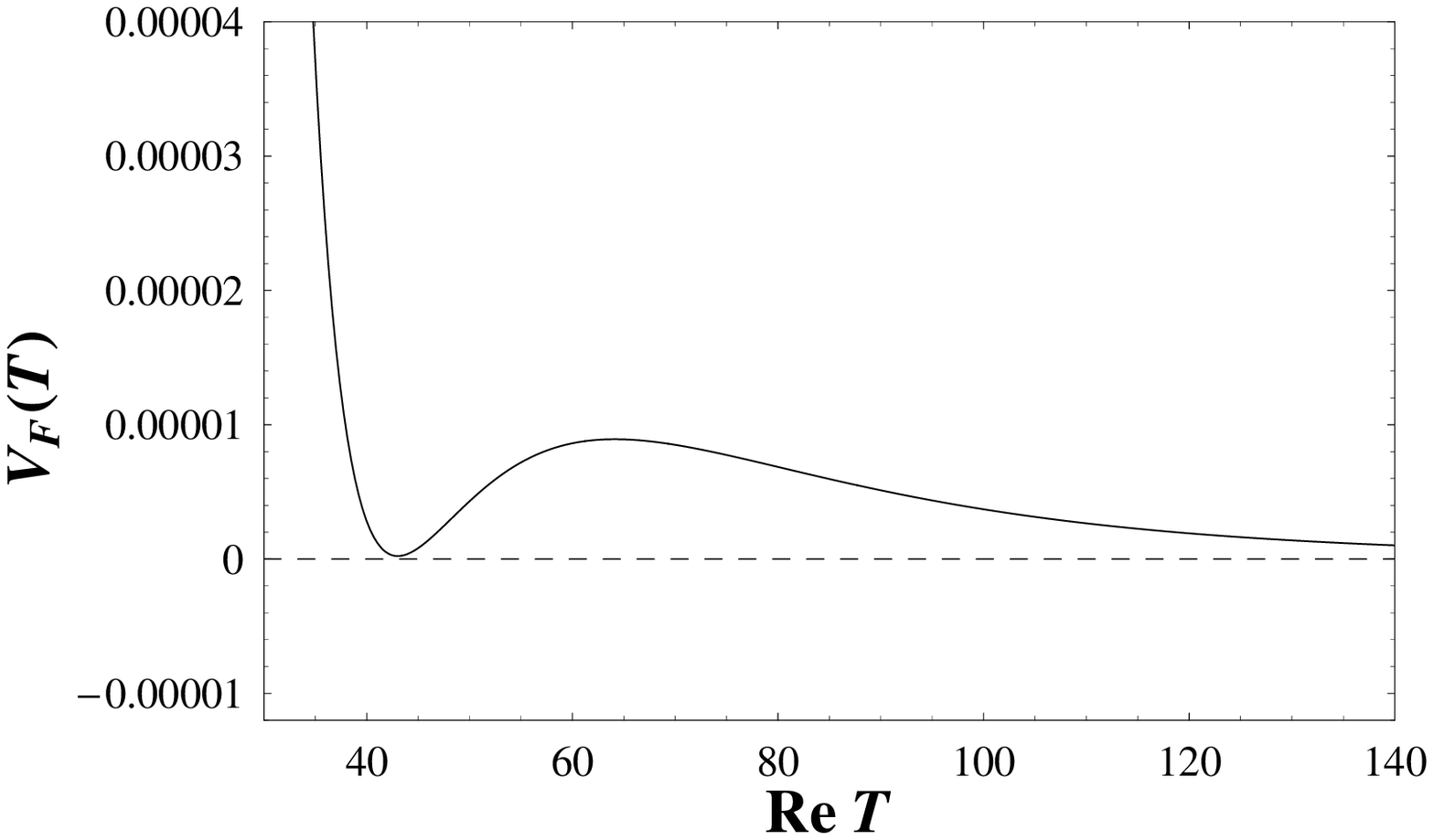,width=12cm} \caption{Solid black:
The F-term scalar potential $V_F(T)$ leading to a dS minimum at
$T\approx 43$ corresponding to a parametrically large volume
$\hat{\cal V}\approx 309$ through the inclusion of the leading
perturbative and non-perturbative effects. The choice or
parameters here reads $\kappa=5$ (as realized on the quintic),
$W_0\approx-32.35$, $a=2\pi/100$ and $\hat{\xi}\approx 7.98$. Note
the smallness of the $\alpha'$-expansion parameter
$\hat{\xi}/(2\hat{\cal V})|_{\rm min}\approx 0.01$ in the minimum.}%
\label{Fig.1}}

Note that the expansion parameter of the $\alpha'$-corrected
K\"ahler potential, $\hat{\xi}/(2\hat{\cal V})$, is invariant
under the rescaling eq.~\eqref{scale1}. It has a value in the
above dS minimum of $\hat{\xi}/(2\hat{\cal V})|_{\rm min}\approx
0.1$ which is already small enough to trust the reliability of the
$\alpha'$-expansion since we neglected higher-order corrections in
$\alpha'$ to $K$.

However, the reliability of the expansion can be thoroughly
improved by noting that the scalar potential eq.~\eqref{VFT}
allows us trade the size of $\hat{\xi}/(2\hat{\cal V})|_{\rm min}$
for the size of $W_0$. Here it is now crucial that, as explained
above, having a large $W_0$ poses no problematic back-reaction in
the type IIB flux compactifications of~\cite{GKP} as there fluxes
must be ISD and of $(1,2)$- and $(0,3)$-type which confines the
back-reaction to the warp factor.

As a demonstration for this we use the explicit example of a
parameter choice $\kappa=5$ (as realized on the quintic),
$W_0\approx-32.35$, $a=2\pi/100$ and $\hat{\xi}\approx 7.98$. This
realizes a dS minimum again at $T\approx 43$ but with an expansion
parameter $\hat{\xi}/(2\hat{\cal V})|_{\rm min}\approx 0.01$ so
small that neglecting the higher-orders in $\alpha'$ should now be
fully justified. In addition, the non-perturbative contribution to
$W$ has a value $\langle A\,\exp(-aT)\rangle\sim 0.05$ which is
small enough to trust the validity of the non-perturbative
superpotential as well. This result is shown in Fig.~\ref{Fig.1}.

\section{The full case - inclusion of $S$ and $U$}\label{STUdS}

The scaling property eq.~\eqref{scale1} used in the examples of
the previous section clearly neglects the structure of the
parameter $\hat{\xi}$ controlling the size of the
$\alpha'$-correction. $\hat{\xi}$ is not just a parameter but it
depends on both $\chi$, the Euler number of the Calabi-Yau, and
the value of the dilaton since we have
$\hat{\xi}\sim-\chi\,(S+\bar{S})^{3/2}$. Thus, if we speak of
rescaling $\hat{\xi}$ this means stabilizing the dilaton at an
appropriate value of ${\rm Re}\,S$ since we cannot rescale
continuously the discrete, topological quantity $\chi$.

\subsection{Stabilizing $S$ and $T$ on the quintic $C\mathbb{P}^4_{1,1,1,1,1}$ explicitly}

Hence, in order to give a realistic example we shall have to
include the stabilization of the dilaton by the fluxes explicitly.
The K\"ahler potential now reads \beq K=-2\cdot \ln\left({\cal
\hat{V}}+\frac{\xi}{2}\,(S+\bar{S})^{3/2}\right)-\ln(S+\bar{S})\;\;.\eeq
The fluxes induce the GVW superpotential eq.~\eqref{fluxW} which
stabilizes $S$ and the complex structure moduli $U^\alpha$.. Its
only dependence on $S$ originates in $G_{(3)}=F_{(3)}-S\cdot
H_{(3)}$ since the integrals $\int_{\rm CY_3}F_{(3)}\wedge\Omega$
and $\int_{\rm CY_3}H_{(3)}\wedge\Omega$ in $W_0$ are determined
entirely by the periods of the Calabi-Yau and thus depend only on
the $U^\alpha$. After integrating out the $U^\alpha$ we can
therefore write the superpotential $W_0$ with respect to $S$
without loss of generality as \beq W_0=C_1-C_2\cdot
S\label{WS}\eeq where we have defined the ($U^\alpha$-dependent)
constants $C_1=\frac{1}{2\pi}\,\int_{\rm CY_3}F_{(3)}\wedge\Omega$
and $C_2=\frac{1}{2\pi}\,\int_{\rm CY_3}H_{(3)}\wedge\Omega$.
Integrating out the complex structure moduli is justified at this
stage as the next subsection demonstrates that they get masses
which are parametrically larger than $S$ and $T$.

In absence of the $\alpha'$-correction the supersymmetric
stationary point for $S$ is given from $D_SW=0$ as \beq
S_0=-\frac{C_1+\langle A\,e^{-aT}\rangle}{C_2}\;\;.\label{S0}\eeq
Since typically it is $\langle A\,e^{-aT}\rangle\sim 0.1$, $S_0$
is determined completely by the flux constants if $|C_1|,|C_2|>1$.

Turning on the $\alpha'$-correction we expect that the true value
of the minimum for the dilaton is close by the unperturbed SUSY
point, $S_{\rm min}\approx S_0$, if we choose $|C_1|,|C_2|>1$ such
that $S_0\gg 1$.

As a test this expectation we will now display the complete
$S$-$T$-system for the 4 real fields (2 moduli, 2 axions)
contained in $T=T_r+i\tau$ and $S=S_r+i\sigma$. The scalar
potential for $S$ and $T$ is then given by eq.~\eqref{VFST} as
\bea
V_F(S,T)&=&e^K\,\left\{K^{T\bar{T}}[W_T\overline{W_T}+(W_T\cdot\overline{W
K_T}+c.c.)]\right.\nonumber\\
&&\;\left.+[K^{T\bar{S}}D_TW\overline{D_SW}+c.c]+K^{S\bar{S}}|D_SW|^2\right.\nonumber\\
&&\left.+3\hat{\xi}\,\frac{\hat{\xi}^2+7\hat{\xi}\hat{\cal
V}+\hat{\cal V}^2}{(\hat{\cal V}-\hat{\xi})(\hat{\xi}+2\hat{\cal
V})^2}\,|W|^2\right\}\label{VFST2}\eea

For the sake of explicitness we will take parameters which derive
from the Calabi-Yau given by the quintic
$C\mathbb{P}^4_{1,1,1,1,1}$ which is given by the vanishing locus
of the polynomial \beq z_1^5+z_2^5+z_3^5+z_4^5+z_5^5=0\;\;. \eeq
This threefold has $h^{1,1}=1$, $\kappa=5$ and $\chi=-200$. An
orientifold with $O3$- and $O7$-planes can be formed from this
manifold by using, e.g., the projection~\cite{ibrunner} \beq {\cal
O}=(-1)^{F_L}\Omega_P\,\sigma^\ast\;\; {\rm with}\;\;\sigma
:\;\{z_1,z_2,z_3,z_4,z_5\}\to \{z_2,z_1,z_3,z_4,z_5\}\eeq where
the holomorphic involution $\sigma$ acts on the holomorphic 3-form
$\Omega$ as $\sigma^\ast\Omega=-\Omega$. For this Calabi-Yau we
get $\xi=0.17133$ and we will further use the choice of flux
parameters $C_1=-13.743$ and $C_2=1.4$ as well as $A=1$ and
$a=2\pi/100$ for the non-perturbative sector. For these values we
expect then the minimum for $S$ to be close to $S_0\approx 9.8$.

\FIGURE[ht]{\epsfig{file=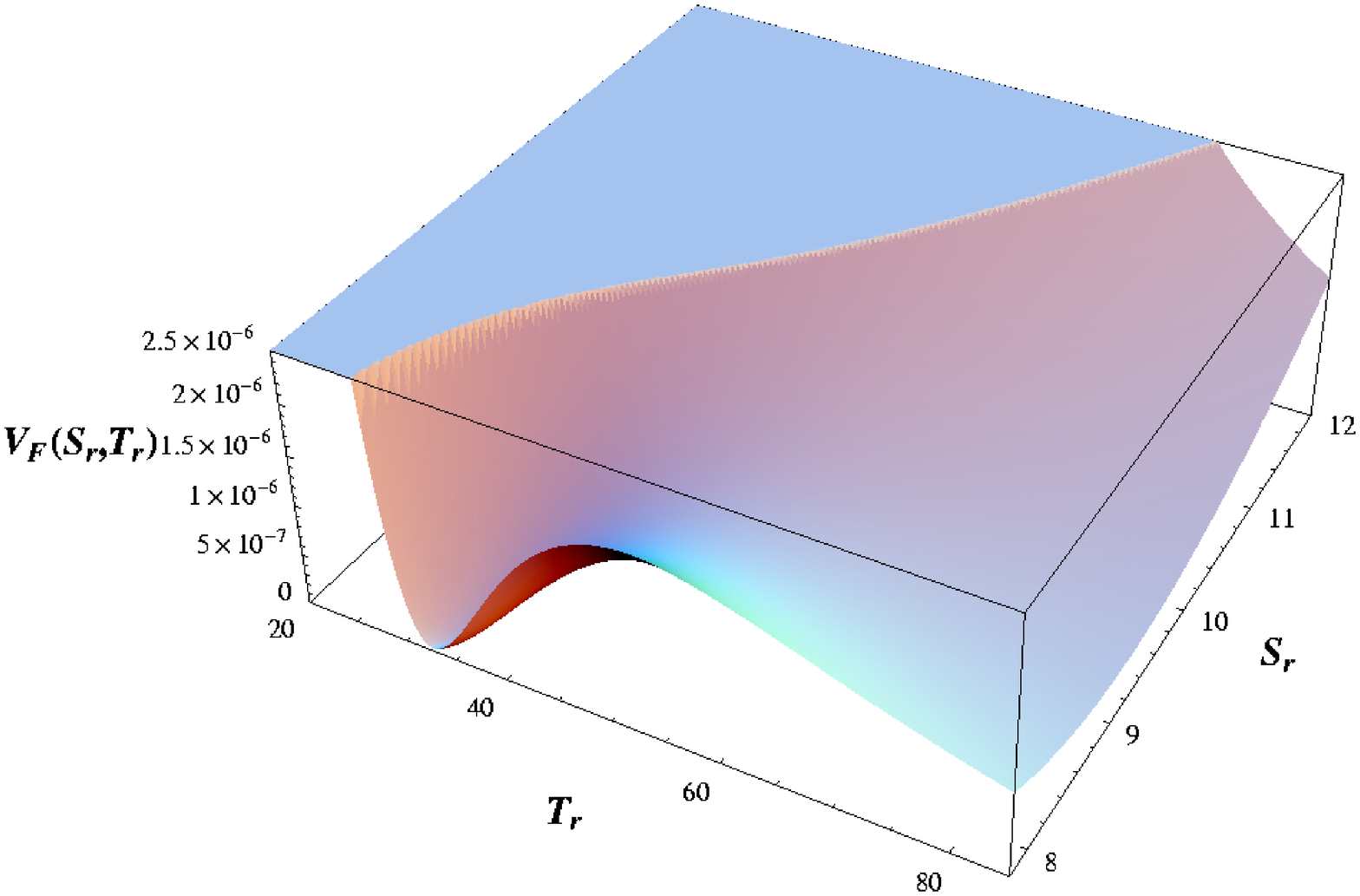,width=7.2cm}
\epsfig{file=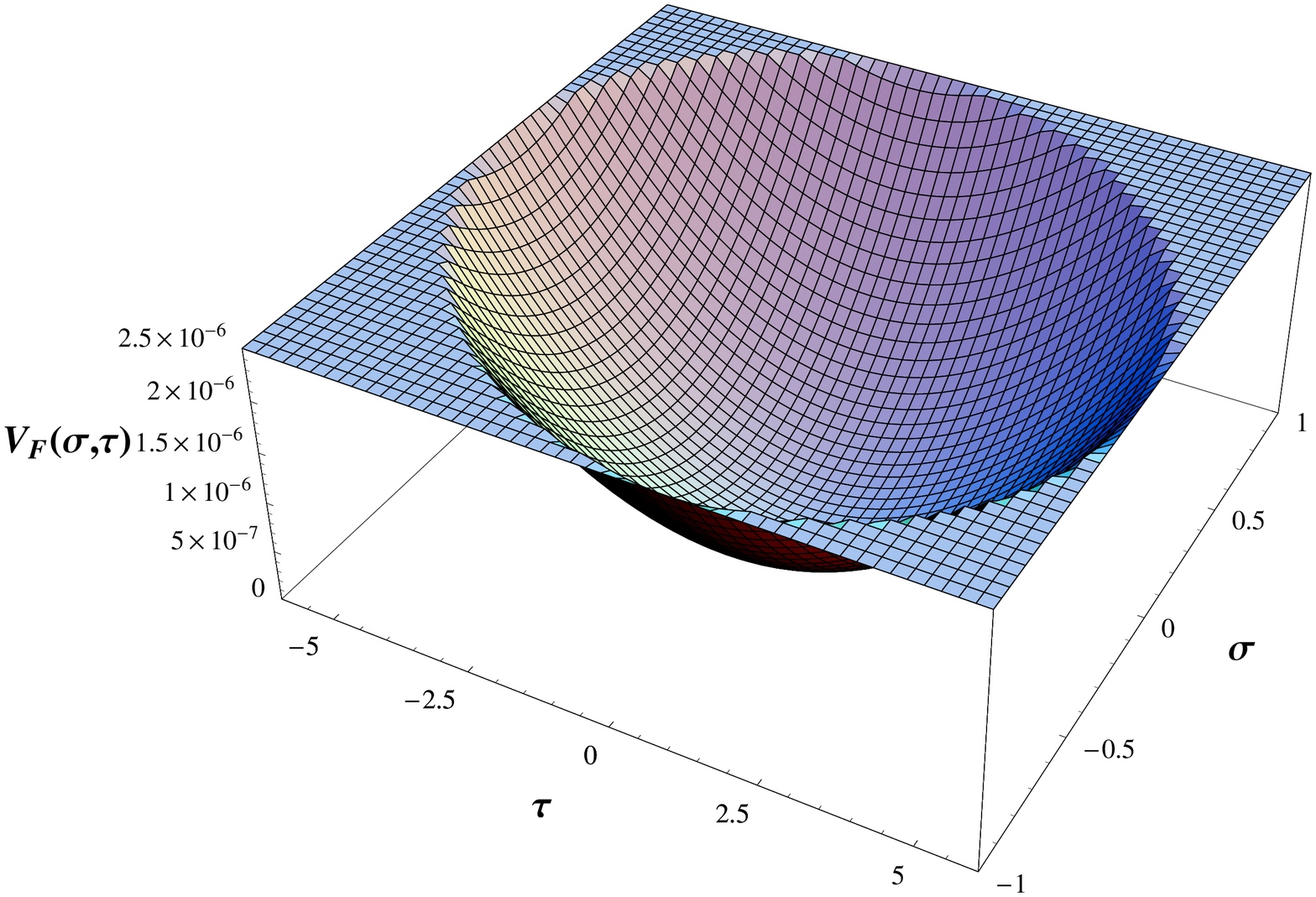,width=7.2cm} \caption{The F-term scalar
potential $V_F(S,T)$ leading to a dS minimum at $T\approx 33.3$
and $S\approx 7.9$ corresponding to a parametrically large volume
$\hat{\cal V}\approx 211$ and weak string coupling $g_S\approx
0.1$ through the inclusion of the leading perturbative and
non-perturbative effects. The choice of the Calabi-Yau, the
quintic, gives the parameters $\kappa=5$, $\chi=-200$ and thus
$\xi=0.17133$. Further it is chosen $a=2\pi/100$, $A=1$ and
$C_1=-13.743$ and $C_2=1.4$. Note the smallness of the
$\alpha'$-expansion parameter
$\hat{\xi}/(2\hat{\cal V})|_{\rm min}\approx 0.03$ in the minimum.}%
\label{Fig.2}}

An analysis of the model yields a dS minimum for all 4 real
scalars at \beq
T_r\approx33.3\;,\;S_r\approx7.9\;,\;\tau=\sigma=0\eeq and thus at
weak string coupling $g_S\approx 0.1$ and large volume $\hat{\cal
V}\approx 211$. The fact, that this stationary point of the scalar
potential is a true minimum one can see at the eigenvalues of the
full $4\times 4$ mass matrix $\partial_m\partial_n
V_F(T_r,S_r,\tau,\sigma)$ (with $m,n\in{T_r,S_r,\tau,\sigma}$)
which are all positive \beq m_{S_r}^2\approx
10^{-5}\;,\;m_{\sigma}^2\approx5\cdot 10^{-6}\;,\;
m_{T_r}^2\approx 6\cdot 10^{-8}\;,\;m_{\tau}^2\approx 1.4\cdot
10^{-7}\;\;.\eeq The value of $S$ in the minimum deviates from its
value in the SUSY stationary point $S_0$ by only 20\% which
justifies the above expectation. Hence, using $D_SW=0$ remains a
good approximation to determine the vacuum value of $S$.

Note that there is a mass hierarchy of ${\cal O}(10^{-2})$ between
the $T$-modulus and the axio-dilaton which is typical for all
cases where $S$ is fixed by fluxes while $T$ is stabilized at
large values using non-perturbative effects.

The structure of this $S$- and $T$-stabilizing dS minimum is
displayed in Fig.~\ref{Fig.2}. Finally, let us calculate the
F-terms in the 2 sectors of the model and the gravitino mass to
get a feeling for the supersymmetry breaking occurring here. We
have \beq m_{3/2}=\langle e^{K/2}|W|\rangle\approx 3\cdot
10^{-2}\eeq and for the F-terms we get\beq F_T=\langle
e^{K/2}D_TW\rangle\approx 10^{-3}\;,\;F_S=\langle
e^{K/2}D_SW\rangle\approx 3\cdot 10^{-4}\;\;.\eeq Thus,
supersymmetry is predominantly broken in the $T$-sector which fits
into the former result that the minimum for $S$ is nearly
supersymmetric. Let us note here the high supersymmetry breaking
scale with a gravitino mass of the order of the GUT scale. This is
a generic feature of the construction since there is no
suppression of the magnitude of $W_0$ as compared with the KKLT
construction. In addition, lowering $m_{3/2}$ would require tuning
$\hat{\cal V}$ to ever larger values which is done by tuning $a$
smaller and $S$ larger - by fluxes which is, however, limited by
the upper bound the flux size in type IIB.

Note further, that here the combination of perturbative and
non-perturbative effects succeed to stabilize $S$ and $T$ in a
true minimum without the presence of any complex structure
modulus. This is different from the case studied, e.g.,
in~\cite{nilles} where upon using only non-perturbative effects
tachyonic directions were found in the absence of complex
structure moduli (for a more recent argument in favor of the
generic stability of the KKLT vacua see~\cite{hebruss}).

\subsection{Complex structure}

We shall now shortly discuss the inclusion of the complex
structure moduli (of which the quintic example contains 101). The
crucial point here is to note that the flux stabilization of the
$U^\alpha$-fields will not be significantly influenced by the
$\alpha'$-correction of the $S$-$T$-sector since the inverse of
the K\"ahler metric decouples the $U^\alpha$ from $S$ and $T$.
From \beq K=\underbrace{-2\ln(\hat{\cal
V}+\hat{\xi}/2)-\ln(S+\bar{S})}_{K_1}\underbrace{-\ln(-i\int_{\rm
CY_3}\bar{\Omega}\wedge\Omega)}_{K_{c.s.}}\Rightarrow
K_{a\alpha}=0\eeq we see immediately that the inverse of the
K\"ahler metric must be block diagonal and thus of the form \beq
K^{A\bar{B}}=\left(\begin{array}{cc} K_1^{a\bar{b}} & 0\\ 0 &
K_{c.s.}^{\alpha\bar{\beta}}\end{array}\right)\eeq where
$K_1^{a\bar{b}}$ is given by eq.~\eqref{K1ijinv}. The only place
where the $S$-$T$-sector can influence the complex structure
moduli is through $W$ itself. However, $W$ appears suppressed by
$K_{U^\alpha}$ in the supercovariant derivative $D_\alpha W$ and
thus for sufficiently large ${\rm Re}\,U^\alpha$ the solution to
$D_\alpha W=0$ should give an excellent approximation to the true
minimum of $U^\alpha$.

For the sake of explicitness we take now the simplified case of
one complex structure modulus $U=U_r+i\cdot\nu$. Then we have (see
e.g.~\cite{LuestW}) \beq K_{c.s.}=-\ln(U+\bar{U})\eeq and it
remains to specify the flux superpotential for $U$. Since this is
not known for the example of the quintic, we take guidance in
toroidal orientifold examples of L\"ust et al.~\cite{LuestW} where
for those with just one complex structure modulus one can write
$W_0$ as \beq W_0=c_1+d_1\cdot U-(c_2+d_2\cdot U)\cdot
S\;\;.\label{WSU}\eeq Here $c_1,d_1,c_2,d_2$ are now true
constants determined entirely by topological information of the
Calabi-Yau.

\FIGURE[ht]{\epsfig{file=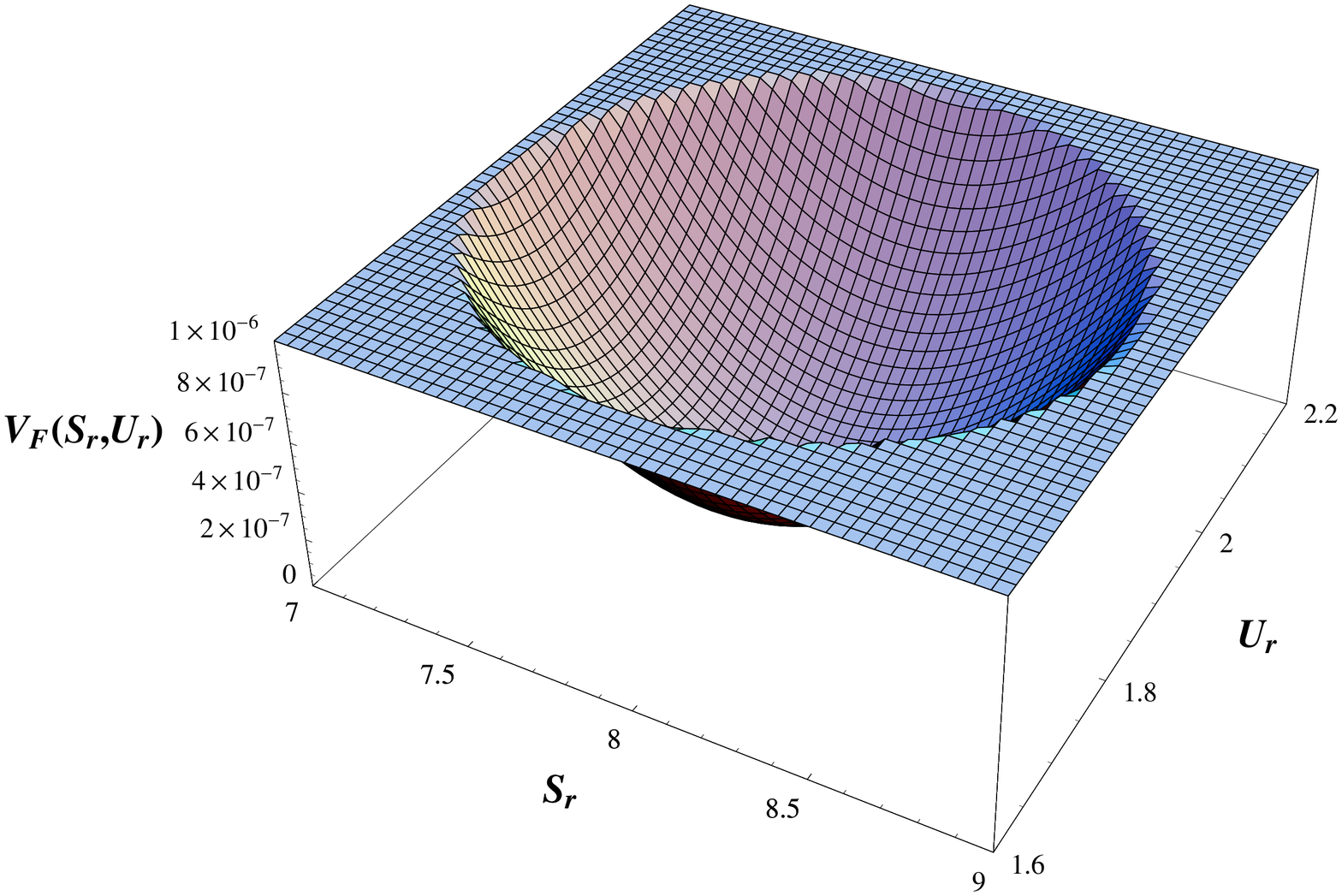,width=12cm} \caption{The
F-term scalar potential $V_F(S,U)$ at $T=T_{\rm min}\approx 32.8$
leading to a dS minimum at $S\approx 7.9$ and $U\approx 1.9$. The
flux parameters are chosen to be $c_1=-6.93$, $c_2=0.7$,
$d_1=-3.46$, $d_2=0.35$ while all other parameters
remain as they were chosen for the quintic in the previous subsection.}%
\label{Fig.3}}

The inverse K\"ahler metric is now \beq
K^{A\bar{B}}=\left(\begin{array}{cc} K_1^{a\bar{b}} & 0\\ 0 &
(U+\bar{U})^2\end{array}\right)\eeq and we get the scalar
potential \bea
V_F(S,T,U)&=&e^K\,\left\{K^{T\bar{T}}[W_T\overline{W_T}+(W_T\cdot\overline{W
K_T}+c.c.)]\right.\nonumber\\
&&\;\left.+[K^{T\bar{S}}D_TW\overline{D_SW}+c.c]+K^{S\bar{S}}|D_SW|^2
+K^{U\bar{U}}|D_UW|^2\right.\nonumber\\
&&\left.+3\hat{\xi}\,\frac{\hat{\xi}^2+7\hat{\xi}\hat{\cal
V}+\hat{\cal V}^2}{(\hat{\cal V}-\hat{\xi})(\hat{\xi}+2\hat{\cal
V})^2}\,|W|^2\right\}\;\;.\label{VFSTU}\eea The supersymmetric
stationary points for $S$ and $U$ are now given by \beq
S_0=-\,\frac{c_1+d_1\,U_0}{c_2+d_2\,U_0}=-\,\frac{d_1}{d_2}\;,\;
U_0=\frac{c_1}{d_1}=\frac{c_2}{d_2}\;\;.\eeq Here we neglected the
contribution $\langle A\exp(-aT)\rangle\sim 0.1$ do to its
smallness. Guided by this, a choice of flux parameters \beq
c_1=-6.93\;,\;c_2=0.7\;,\;d_1=-3.46\;,\;d_2=0.35\eeq should again
lead to $S\approx S_0=9.8$ and now also $U\approx U_0=2$ while
giving a $W_0$ of the same size as in the previous subsection for
the above values of $S$ and $U$. In using the fractional numbers
for the constants $c_1,c_2,d_1,d_2$ we were borrowing from the
fact that in the many complex structure moduli case (as realized
on the quintic) the 3-form flux supported on the many different
3-cycles allows for tuning the constants the same way as the total
$W_0$ is tuned to get the cosmological constant small.

The analysis of the scalar potential eq.~\eqref{VFSTU} reveals a
true minimum at \beq T_r\approx 32.8\;,\;S_r\approx
7.9\;,\;U_r\approx 1.9\;\;.\eeq This stationary point is again a
minimum as seen from the masses\bea m_{U_r}^2&\approx&3\cdot
10^{-5}\;,\; m_{S_r}^2\approx3\cdot
10^{-6}\;,\;m_{T_r}^2\approx1.5\cdot 10^{-8}\nonumber\\
m_{\nu}^2&\approx&3\cdot 10^{-5}\;,\; m_{\sigma}^2\approx
10^{-6}\;,\;m_{\tau}^2\approx4\cdot 10^{-8}\;\;.\eea It is de
Sitter and looks practically unchanged with respect to $S$ and
$T$. The axions of $T$ and $S$ remain stabilized at
$\tau=\sigma=0$ as before and it is clear that the axion of $U$ is
fixed at $\nu=0$ the same way as $\sigma$ since they enter $W$ in
the same way. Fig.~\ref{Fig.3} displays the minimum for the real
parts of $S$ and $U$.

A property of both the $S$-$T$-$U$-model presented here and the
$S$-$T$-model of the last subsection is that the volume modulus
$T_r$ remains to be the lightest field after stabilization.

Note that $U_r$ deviates by less than 5\% from its supersymmetric
stationary point. This fits with the result for the F-terms in the
different sectors \beq F_T\approx 7\cdot 10^{-4}\;,\;F_S\approx
2\cdot 10^{-4}\;,\;F_U\approx 9\cdot 10^{-5}\eeq which again shows
the dominance of the $T$-sector SUSY breaking as well as that SUSY
is most weakly broken in the $U$-sector. The gravitino mass is
\beq m_{3/2}\approx 10^{-2}\;\;.\eeq

The smallness of the deviation which $U$ has compared with its
SUSY stationary point makes it possible to check the numerical
results with an analytic expansion of the solution for $U$ away
from the SUSY point $U_0$ in terms of powers of
$\hat{\xi}/(2\hat{\cal V})\approx 0.03$. We use here a method
developed in~\cite{BinDud96} which works as follows: Take the
scalar potential for $U$, plug in an expansion
$U=U_0\cdot(1+a_1\frac{\hat{\xi}}{2\hat{\cal V}}+\ldots)$ and
determine the perturbed stationary point in $U$ by determining the
coefficient $a_1$ from $\partial_{a_1} V_F(U)=0$. Following this
recipe we arrive at an expression for the first expansion
coefficient \beq a_1=-\left\langle
4A\,e^{-aT_r}\gamma\,\left(\frac{T_r}{S_r}\right)^{3/2}\cdot {\cal
O}(1)\right\rangle\approx -1.67\;\;.\eeq If we plug this result
back into the expansion of $U$ around $U_0$ we get \beq
U=U_0\left(1+a_1\frac{\hat{\xi}}{2\hat{\cal
V}}\right)\approx1.91\eeq which agrees with the full numerical
solution up to 1\% and thus shows the reliability of the numerical
result.

We discussed here the case of one complex structure modulus
explicitly. However, the example used here, the Calabi-Yau given
by the quintic hypersurface $C\mathbb{P}^4_{1,1,1,1,1}$, has
$h^{2,1}=101$ complex structure moduli. The stabilization of all
of them cannot be done explicitly in a practically feasible way.
However, since all complex structure moduli enter the flux
superpotential in the qualitatively the same way as the first one
discussed above, and for this first one the SUSY condition
$D_UW=0$ gives a rather good approximation for its minimum, the
stabilization of several complex structure moduli should proceed
without major changes along the lines of~\cite{GKP}.

Hence, it seems to be reasonably safe to conclude that this
construction stabilizes all geometric moduli in a candidate string
example given by $C\mathbb{P}^4_{1,1,1,1,1}$ explicitly in a
tunable dS vacuum at parametrically large volume and weak string
coupling.

\subsection{Application to all smooth hypersurfaces in
$C\mathbb{P}^4$ with $h^{1,1}=1$}\label{Ex}

We shall now try to extend the results of the previous section to
other Calabi-Yau 3-folds than $C\mathbb{P}^4_{1,1,1,1,1}$. To
carry over as much of the structure as possible, we restrict
ourselves to the case $h^{1,1}=1$ although there are no reasons of
principle which should prevent us from extending the construction
to $h^{1,1}>1$, too (for instance, it should be rather
straightforward to apply our construction to the
$C\mathbb{P}^4_{1,1,1,6,9}$-model of~\cite{explicitmodels}). There
are three other non-singular hypersurfaces in projective 4-space
$C\mathbb{R}^4$ which yield a Calabi-Yau 3-fold with $h^{1,1}=1$
besides $C\mathbb{P}^4_{1,1,1,1,1}$. We list in Table~\ref{Tab.1}
the four manifolds and their relevant properties for
comparison~\cite{KlTh}. \TABLE[t]{
\begin{tabular}{lllll}
& $C\mathbb{P}^4_{1,1,1,1,1}$ & $C\mathbb{P}^4_{2,1,1,1,1}$ & $C\mathbb{P}^4_{4,1,1,1,1}$ & $C\mathbb{P}^4_{5,2,1,1,1}$\\
\hline
$h^{2,1}$ &  101 & 103  & 149  &  145  \\
$\chi$ & -200  & -204  & -296  &  -288  \\
$\kappa$ & 5  & 3  & 2  & 1   \\
\end{tabular}
\caption{quintic hypersurfaces in $C\mathbb{P}^4$} \label{Tab.1}}

From this synopsis it is clear that all four Calabi-Yau 3-folds
satisfy the necessary requirement $h^{2,1}>h^{1,1}$ to be eligible
for the procedure of moduli stabilization with K\"ahler uplifting
from the preceding sections. In particular, the results for
$C\mathbb{P}^4_{1,1,1,1,1}$ should carry over nearly unchanged to
$C\mathbb{P}^4_{2,1,1,1,1}$ as there $\chi$ is practically the
same. The smaller value of the self intersection $\kappa$ on
$C\mathbb{P}^4_{2,1,1,1,1}$ will require us to stabilize the
dilaton at even weaker string coupling compared to the former case
of $C\mathbb{P}^4_{1,1,1,1,1}$ in order to maintain the same
minimum for $T$. The other two possibilities will require
stabilizing the dilaton at a relatively stronger string coupling
as there the Euler numbers are about 40-50\% larger in size.

\section{Conclusion}\label{con}

In this paper we discussed (building on an earlier analysis
by~\cite{Brama}) de Sitter vacua in string theory which arise by
spontaneously breaking supersymmetry through F-terms induced by
the leading quantum correction to the K\"ahler potential of the
K\"ahler moduli. This correction arises from the known $R^4$-term
of type IIB string theory at ${\cal O}(\alpha'^3)$. Taking into
account the dependence of this correction on the dilaton allows us
to stabilize both the volume K\"ahler modulus and the dilaton in a
metastable dS minimum which is at parametrically large volume of
${\cal O}(100)$ and weak string coupling $g_S\sim 0.1$. In
addition, we are able to take the interplay between the leading
non-perturbative contribution to the superpotential and the
leading perturbative K\"ahler correction into a regime where both
are ${\cal O}(0.1)$ or even smaller. Hence, these dS minima are
reliable as both the perturbative $\alpha'$-expansion and the
non-perturbative expansion in the superpotential are under
parametrical control. It is crucial here, that in type IIB the
back-reaction of fluxes of $(1,2)$- and $(0,3)$-type is confined
to changes in the warp factor. This, in turn allows us to make the
$\alpha'$-expansion parameter small by trading its size for that
of the flux superpotential.

Let us note here, that the bound on the size of 3-form flux and
thus the size of $W_0$ in type IIB~\cite{vacstat} may prevent us
from getting exponentially large volumes: Should the limit on flux
size also yield a lower bound of $g_S$, then this limits the
maximal rank of the gauge groups in the non-perturbative sector
usable for tuning the volume to large values. The gauge group
rank, however, sets the scale of the achievable volume, which thus
can be exponentially large only if the rank can be made very large
and $g_S$ very small.

As the vacuum value of the superpotential in these vacua is
generically ${\cal O}(1\ldots 10)$, any non-exponentially small
lower bound on $g_S$ in type IIB flux compactifications would
imply a necessarily high-scale gravitino mass and supersymmetry
breaking scale. In the examples discussed, we get typical values
of $m_{3/2}\sim 10^{-3}\ldots 10^{-2}$.

In the next step we included the stabilization of a single complex
structure modulus by fluxes explicitly. Since it turns out that
this modulus remains stabilized at values very close to the one
dictated by the supersymmetry condition in the complex structure
moduli sector, this implies that the stabilization of several
complex structure moduli is straight forward along the general
procedure of type IIB flux compactifications. This feature allows
us to conclude that the vacua constructed apply directly to a
sub-class of type IIB flux compactifications on Calabi-Yau 3-folds
with $h^{1,1}=1$ which is given by the quintic hypersurfaces in
projective 4-space $C\mathbb{P}^4$. Extending the construction to
the case $h^{1,1}>1$ should be straightforward.

All four members of this class ($C\mathbb{P}^4_{1,1,1,1,1}$,
$C\mathbb{P}^4_{2,1,1,1,1}$, $C\mathbb{P}^4_{4,1,1,1,1}$,
$C\mathbb{P}^4_{5,2,1,1,1}$) satisfy the requirements of the
construction, namely that $h^{2,1}>h^{1,1}$. Furthermore, the
effects used are reasonably sound effects of string theory
corroborated by solid string calculations (especially the ${\cal
O}(\alpha'^3)$-correction). Viewed together, this gives an
explicit construction of dS vacua in string theory which stabilize
all geometric moduli in these four examples.

Interesting cosmological questions arise now in these models.
Since the cosmological constant of these dS vacua is amenable to
fine-tuning by the fluxes, one may want to look for realizations
of inflation in this scenario. Here one may revisit, e.g., the
mechanism of inflation driven by open string
$D3$-$\overline{D3}$-distance modulus~\cite{KKLMMT} or axionic
directions of the closed string moduli potential along the lines
of~\cite{blanco,West,betterracetinf,huta}, which is however left
for future work.

Let us finally mention, that it may be useful to revisit the
question of stabilizing the dilaton by a combination of
$H_{(3)}$-flux and gaugino condensation in the heterotic
string~\cite{hetcond}, since there the $R^4$-term corrects the
K\"ahler potential of the heterotic dilaton~\cite{R4het}. As in
the models discussed here, there is no need to have $|W_0|$ small
while we get large values for $T$. This could translate in the
heterotic case into a viable way to stabilize the dilaton at the
phenomenologically required value $S\sim 2$ with having a flux
superpotential of ${\cal O}(1)$ as required there by flux
quantization.

\acknowledgments

I am particularly indebted to R.~Kallosh for numerous intensive
discussions at different stages throughout this work. I would like
to thank A.~Hebecker, F.~Quevedo and M.~Serone for useful
discussions and comments.


\begin{thebibliography}{999}
\bibitem{BoussoP}
R.~Bousso \& J.~Polchinski, JHEP {\bf 0006}, 006 (2000)
[arXiv:\hepth{0004134}].
%%CITATION = HEP-TH 0004134;%%

\bibitem{kklt}
S.~Kachru, R.~Kallosh, A.~Linde \& S.~P.~Trivedi, Phys.\ Rev.\ D
{\bf 68}, 046005 (2003) [arXiv:\hepth{0301240}].
%%CITATION = HEP-TH 0301240;%%

\bibitem{sussk}
L.~Susskind, [arXiv:\hepth{0302219}].
%%CITATION = HEP-TH 0302219;%%

\bibitem{dougl}
M.~R.~Douglas, JHEP {\bf 0305}, 046 (2003)
[arXiv:\hepth{0303194}].
%%CITATION = HEP-TH 0303194;%%

\bibitem{WMAP3}
D.~N.~Spergel {\it et al.}, \emph{submitted to Astrophys. J.},
[arXiv:astro-ph/0603449].
%%CITATION = ASTRO-PH 0603449;%%

\bibitem{GKP}
S.~B.~Giddings, S.~Kachru \& J.~Polchinski, Phys.\ Rev.\ D {\bf
66}, 106006 (2002) [arXiv:\hepth{0105097}].
%%CITATION = HEP-TH 0105097;%%

\bibitem{CBachas}
C.~Bachas, [arXiv:\hepth{9503030}].
%%CITATION = HEP-TH 9503030;%%

\bibitem{PolStrom}
J.~Polchinski \& A.~Strominger, Phys.\ Lett.\ B {\bf 388} (1996)
736 [arXiv:\hepth{9510227}].
%%CITATION = HEP-TH 9510227;%%

\bibitem{Michelson}
J.~Michelson, Nucl.\ Phys.\ B {\bf 495} (1997) 127
[arXiv:\hepth{9610151}].
%%CITATION = HEP-TH 9610151;%%

\bibitem{DasSeRa}
K.~Dasgupta, G.~Rajesh \& S.~Sethi, JHEP {\bf 9908} (1999) 023
[arXiv:\hepth{9908088}].
%%CITATION = HEP-TH 9908088;%%

\bibitem{TaylVaf}
T.~R.~Taylor \& C.~Vafa, Phys.\ Lett.\ B {\bf 474} (2000) 130
[arXiv:\hepth{9912152}].
%%CITATION = HEP-TH 9912152;%%

\bibitem{GVW}
S.~Gukov, C.~Vafa \& E.~Witten, Nucl.\ Phys.\ B {\bf 584}, 69
(2000) [Erratum-ibid.\ B {\bf 608}, 477 (2001)]
[arXiv:\hepth{9906070}].
%%CITATION = HEP-TH 9906070;%%

\bibitem{Vafa}
C.~Vafa, J.\ Math.\ Phys.\  {\bf 42}, 2798 (2001)
[arXiv:\hepth{0008142}].
%%CITATION = HEP-TH 0008142;%%

\bibitem{Mayr}
P.~Mayr, Nucl.\ Phys.\ B {\bf 593} (2001) 99
[arXiv:\hepth{0003198}].
%%CITATION = HEP-TH 0003198;%%

\bibitem{GSS}
B.~R.~Greene, K.~Schalm \& G.~Shiu, Nucl.\ Phys.\ B {\bf 584}
(2000) 480
\\{}[arXiv:\hepth{0004103}].
%%CITATION = HEP-TH 0004103;%%

\bibitem{KlebStrass}
I.~R.~Klebanov \& M.~J.~Strassler, JHEP {\bf 0008}, 052 (2000)
[arXiv:\hepth{0007191}].
%%CITATION = HEP-TH 0007191;%%

\bibitem{Curio2}
G.~Curio \& A.~Krause, Nucl.\ Phys.\ B {\bf 602}, 172 (2001)
[arXiv:\hepth{0012152}].
%%CITATION = HEP-TH 0012152;%%

\bibitem{CKLT}
G.~Curio, A.~Klemm, D.~L{\"u}st \& S.~Theisen, Nucl.\ Phys.\ B
{\bf 609} (2001) 3
\\{}[arXiv:\hepth{0012213}];\\
%%CITATION = HEP-TH 0012213;%%
G.~Curio, A.~Klemm, B.~K{\"o}rs \& D.~L{\"u}st, Nucl.\ Phys.\ B
{\bf 620} (2002) 237
\\{}[arXiv:\hepth{0106155}].
%%CITATION = HEP-TH 0106155;%%

\bibitem{HaaLou}
M.~Haack \& J.~Louis, Nucl.\ Phys.\ B {\bf 575} (2000) 107
[arXiv:\hepth{9912181}]; \\
%%CITATION = HEP-TH 9912181;%%
Phys.\ Lett.\ B {\bf 507} (2001) 296 [arXiv:\hepth{0103068}].
%%CITATION = HEP-TH 0103068;%%

\bibitem{BB}
K.~Becker \& M.~Becker, JHEP {\bf 0107} (2001) 038
[arXiv:\hepth{0107044}].
%%CITATION = HEP-TH 0107044;%%

\bibitem{DallAgata}
G.~Dall'Agata, JHEP {\bf 0111} (2001) 005 [arXiv:\hepth{0107264}].
%%CITATION = HEP-TH 0107264;%%

\bibitem{KaScTr}
S.~Kachru, M.~B.~Schulz \& S.~Trivedi, JHEP {\bf 0310} (2003) 007
[arXiv:\hepth{0201028}].
%%CITATION = HEP-TH 0201028;%%

\bibitem{silver}
E.~Silverstein, [arXiv:\hepth{0106209}];\\
%%CITATION = HEP-TH 0106209;%%
A.~Maloney, E.~Silverstein \& A.~Strominger,
[arXiv:\hepth{0205316}].
%%CITATION = HEP-TH 0205316;%%

\bibitem{acharya}
B.~S.~Acharya, [arXiv:\hepth{0212294}].
%%CITATION = HEP-TH 0212294;%%

\bibitem{dlust}
R.~Blumenhagen, D.~L\"ust \& T.~R.~Taylor, Nucl.\ Phys.\ B {\bf
663}, 319 (2003)
[arXiv:\hepth{0303016}].\\
%%CITATION = HEP-TH 0303016;%%
J.~F.~G.~Cascales \& A.~M.~Uranga, JHEP {\bf 0305}, 011 (2003)
[arXiv:\hepth{0303024}].
%%CITATION = HEP-TH 0303024;%%

\bibitem{Verl}
H.~Verlinde, Nucl.\ Phys.\ B {\bf 580}, 264 (2000)
[arXiv:\hepth{9906182}].
%%CITATION = HEP-TH 9906182;%%

\bibitem{Curio1}
G.~Curio \& A.~Krause, Nucl.\ Phys.\ B {\bf 643}, 131 (2002)
[arXiv:\hepth{0108220}].
%%CITATION = HEP-TH 0108220;%%

\bibitem{bkqu}
C.~P.~Burgess, R.~Kallosh \& F.~Quevedo, JHEP {\bf 0310}, 056
(2003) [arXiv:\hepth{0309187}].
%%CITATION = HEP-TH 0309187;%%

\bibitem{JockersLouis}
H.~Jockers \& J.~Louis, Nucl.\ Phys.\ B {\bf 718}, 203 (2005)
[arXiv:\hepth{0502059}]
%%CITATION = HEP-TH 0502059;%%

\bibitem{BDKP}
P.~Binetruy, G.~Dvali, R.~Kallosh \& A.~Van Proeyen, Class.\
Quant.\ Grav.\  {\bf 21}, 3137 (2004) [arXiv:\hepth{0402046}].
%%CITATION = HEP-TH 0402046;%%

\bibitem{Nilles2}
K.~Choi, A.~Falkowski, H.~P.~Nilles \& M.~Olechowski, Nucl.\
Phys.\ B {\bf 718}, 113 (2005) [arXiv:\hepth{0503216}].
%%CITATION = HEP-TH 0503216;%%

\bibitem{DuVe}
E.~Dudas \& S.~K.~Vempati, Nucl.\ Phys.\ B {\bf 727}, 139 (2005)
[arXiv:\hepth{0506172}].
%%CITATION = HEP-TH 0506172;%%

\bibitem{VZ}
G.~Villadoro \& F.~Zwirner, Phys.\ Rev.\ Lett.\  {\bf 95}, 231602
(2005) [arXiv:\hepth{0508167}].
%%CITATION = HEP-TH 0508167;%%

\bibitem{Carlos}
A.~Achucarro, B.~de Carlos, J.~A.~Casas \& L.~Doplicher,
[arXiv:\hepth{0601190}].
%%CITATION = HEP-TH 0601190;%%

\bibitem{ChoiJeong}
K.~Choi \& K.~S.~Jeong, JHEP {\bf 0608}, 007 (2006)
[arXiv:\hepth{0605108}].
%%CITATION = HEP-TH 0605108;%%

\bibitem{DuMamb}
E.~Dudas \& Y.~Mambrini, JHEP {\bf 0610}, 044 (2006)
[arXiv:\hepth{0607077}].
%%CITATION = HEP-TH 0607077;%%

\bibitem{Luest}
M.~Haack, D.~Krefl, D.~Lust, A.~Van Proeyen \& M.~Zagermann,
arXiv:\hepth{0609211}.
%%CITATION = HEP-TH 0609211;%%

\bibitem{hebtrap}
A.~P.~Braun, A.~Hebecker \& M.~Trapletti, [arXiv:\hepth{0611102}].
%%CITATION = HEP-TH 0611102;%%

\bibitem{PW}
S.~L.~Parameswaran \& A.~Westphal, JHEP {\bf 0610}, 079 (2006)
[arXiv:\hepth{0602253}].
%%CITATION = HEP-TH 0602253;%%

\bibitem{HG}
G.~von Gersdorff \& A.~Hebecker, Phys.\ Lett.\ B {\bf 624}, 270
(2005) [arXiv:\hepth{0507131}].
%%CITATION = HEP-TH 0507131;%%

\bibitem{BHK2}
M.~Berg, M.~Haack \& B.~Kors, Phys.\ Rev.\ Lett.\  {\bf 96},
021601 (2006) [arXiv:\hepth{0508171}].
%%CITATION = HEP-TH 0508171;%%

\bibitem{zurab}
F.~Paccetti~Correia, M.~G.~Schmidt \& Z.~Tavartkiladze,
[arXiv:\hepth{0608058}].
%%CITATION = HEP-TH 0608058;%%

\bibitem{BHK1}
M.~Berg, M.~Haack \& B.~Kors, JHEP {\bf 0511}, 030 (2005)
[arXiv:\hepth{0508043}].
%%CITATION = HEP-TH 0508043;%%

\bibitem{SaSilv}
A.~Saltman \& E.~Silverstein, JHEP {\bf 0411}, 066 (2004)
[arXiv:\hepth{0402135}].
%%CITATION = HEP-TH 0402135;%%

\bibitem{NillFterm}
O.~Lebedev, H.~P.~Nilles, \& M.~Ratz, Phys.\ Lett.\ B {\bf 636},
126 (2006) [arXiv:\hepth{0603047}].
%%CITATION = HEP-TH 0603047;%%

\bibitem{Acharya2}
B.~Acharya, K.~Bobkov, G.~Kane, P.~Kumar \& D.~Vaman, Phys.\ Rev.\
Lett.\  {\bf 97}, 191601 (2006) [arXiv:\hepth{0606262}].
%%CITATION = HEP-TH 0606262;%%

\bibitem{ISS}
K.~Intriligator, N.~Seiberg \& D.~Shih, JHEP {\bf 0604}, 021
(2006) [arXiv:\hepth{0602239}].
%%CITATION = HEP-TH 0602239;%%

\bibitem{DuISS}
E.~Dudas, C.~Papineau \& S.~Pokorski, [arXiv:\hepth{0610297}].
%%CITATION = HEP-TH 0610297;%%

\bibitem{AHKO}
H.~Abe, T.~Higaki, T.~Kobayashi \& Y.~Omura,
[arXiv:\hepth{0611024}]
%%CITATION = HEP-TH 0611024;%%

\bibitem{OKKLT1}
F.~Br\"ummer, A.~Hebecker \& M.~Trapletti, Nucl.\ Phys.\ B {\bf
755}, 186 (2006) [arXiv:\hepth{0605232}].
%%CITATION = HEP-TH 0605232;%%

\bibitem{OKKLT2}
R.~Kallosh \& A.~Linde, [arXiv:\hepth{0611183}].
%%CITATION = HEP-TH 0611183;%%

\bibitem{ReinoScrucca}
M.~Gomez-Reino \& C.~A.~Scrucca, JHEP {\bf 0605}, 015 (2006)
[arXiv:\hepth{0602246}].
%%CITATION = HEP-TH 0602246;%%

\bibitem{bbhl}
K.~Becker, M.~Becker, M.~Haack \& J.~Louis, JHEP {\bf 0206}, 060
(2002) [arXiv:\hepth{0204254}].
%%CITATION = HEP-TH 0204254;%%

\bibitem{Brama}
V.~Balasubramanian \& P.~Berglund, JHEP {\bf 0411}, 085 (2004)
[arXiv:\hepth{0408054}].
%%CITATION = HEP-TH 0408054;%%

\bibitem{West}
A.~Westphal, JCAP {\bf 0511}, 003 (2005) [arXiv:\hepth{0507079}].
%%CITATION = HEP-TH 0507079;%%

\bibitem{Bobk}
K.~Bobkov, JHEP {\bf 0505}, 010 (2005) [arXiv:\hepth{0412239}].
%%CITATION = HEP-TH 0412239;%%

\bibitem{deAlwis}
S.~P.~de~Alwis, Phys.\ Lett.\ B {\bf 626}, 223 (2005)
[arXiv:\hepth{0506266}].
%%CITATION = HEP-TH 0506266;%%

\bibitem{explicitmodels}
F.~Denef, M.~R.~Douglas \& B.~Florea, JHEP {\bf 0406} (2004) 034
[arXiv:\hepth{0404257}].
%%CITATION = HEP-TH 0404257;%%

\bibitem{GreenSethi}
M.~B.~Green \& S.~Sethi, Phys.\ Rev.\ D {\bf 59}, 046006 (1999)
[arXiv:\hepth{9808061}].
%%CITATION = HEP-TH 9808061;%%

\bibitem{tensort}
S.~Frolov, I.~R.~Klebanov \& A.~A.~Tseytlin, Nucl.\ Phys.\ B {\bf
620}, 84 (2002) [arXiv:\hepth{0108106}].
%%CITATION = HEP-TH 0108106;%%

\bibitem{Brama2}
V.~Balasubramanian, P.~Berglund, J.~P.~Conlon \& F.~Quevedo, JHEP
{\bf 0503}, 007 (2005) [arXiv:\hepth{0502058}].
%%CITATION = HEP-TH 0502058;%%

\bibitem{ibrunner}
I.~Brunner and K.~Hori, JHEP {\bf 0411}, 005 (2004)
[arXiv:hep-th/0303135];\\{}
I.~Brunner, K.~Hori, K.~Hosomichi and
J.~Walcher, [arXiv:hep-th/0401137].
%%CITATION = HEP-TH 0401137;%%
%%CITATION = HEP-TH 0303135;%%

\bibitem{nilles}
K.~Choi, A.~Falkowski, H.~P.~Nilles, M.~Olechowski \& S.~Pokorski,
JHEP {\bf 0411} (2004) 076 [arXiv:\hepth{0411066}].
%%CITATION = HEP-TH 0411066;%%

\bibitem{hebruss}
A.~Hebecker \& J.~March-Russell, [arXiv:\hepth{0607120}].
%%CITATION = HEP-TH 0607120;%%

\bibitem{LuestW}
D.~Lust, S.~Reffert, W.~Schulgin \& S.~Stieberger,
[arXiv:\hepth{0506090}].
%%CITATION = HEP-TH 0506090;%%

\bibitem{BinDud96}
P.~Binetruy \& E.~Dudas, Phys.\ Lett.\ B {\bf 389}, 503 (1996)
[arXiv:\hepth{9607172}].
%%CITATION = HEP-TH 9607172;%%

\bibitem{KlTh}
A.~Klemm \& S.~Theisen, Nucl.\ Phys.\ B {\bf 389}, 153 (1993)
[arXiv:\hepth{9205041}].
%%CITATION = HEP-TH 9205041;%%

\bibitem{vacstat}
M.~R.~Douglas, JHEP {\bf 0305}, 046 (2003)
[arXiv:\hepth{0303194}].
%%CITATION = HEP-TH 0303194;%%

\bibitem{KKLMMT}
S.~Kachru, R.~Kallosh, A.~Linde, J.~Maldacena, L.~McAllister and
S.~P.~Trivedi, JCAP {\bf 0310}, 013 (2003)
[arXiv:\hepth{0308055}].
%%CITATION = HEP-TH 0308055;%%

\bibitem{blanco}
J.~J.~Blanco-Pillado {\it et al.}, JHEP {\bf 0411}, 063 (2004)
[arXiv:\hepth{0406230}].
%%CITATION = HEP-TH 0406230;%%

\bibitem{betterracetinf}
J.~J.~Blanco-Pillado {\it et al.}, JHEP {\bf 0609}, 002 (2006)
[arXiv:\hepth{0603129}].
%%CITATION = HEP-TH 0603129;%%

\bibitem{huta}
R.~Holman \& J.~A.~Hutasoit, [arXiv:\hepth{0603246}].
%%CITATION = HEP-TH 0603246;%%

\bibitem{hetcond}
M.~Dine, R.~Rohm, N.~Seiberg \& E.~Witten, Phys.\ Lett.\ B {\bf
156}, 55 (1985).
%%CITATION = PHLTA,B156,55;%%

\bibitem{R4het}
L.~Anguelova \& D.~Vaman,  Nucl.\ Phys.\ B {\bf 733}, 132 (2006)
[arXiv:\hepth{0506191}].
%%CITATION = HEP-TH 0506191;%%

\end{thebibliography}
\end{document}